\documentstyle[amssymb,12pt]{article}

\newif\iffigs\figstrue
\figsfalse

\iffigs
  \input epsf
\else
\fi

\batchmode
  \newfont{\footscrfont}{rsfs10}
  \newfont{\footbbbfont}{msbm10}
\errorstopmode

\newif\ifscrf\scrftrue
\ifx\footscrfont\nullfont
  \scrffalse
\fi

\newif\ifamsf\amsftrue
\ifx\footbbbfont\nullfont
  \amsffalse
\fi

\def\ppnumber{\vbox{\baselineskip14pt\hbox{CK-TH-2000-002}
\hbox{hep-th/0001217}}}
\def\ppdate{January 2000}
\def\pplogo{\vbox{\kern-\headheight\kern -15pt
\halign{##&##\hfil\cr&{
\ppnumber}\cr\rule{0pt}{2.5ex}&\ppdate\cr}
}}
\makeatletter
\date{}
\def\dedicatory#1{\def\@date{\normalsize\it#1}}
\def\subjclass#1{\def\@thefnmark{}\@footnotetext{1991
    {\it Mathematics Subject Classification.} #1}}
\def\keywords#1{\def\@thefnmark{}\@footnotetext{
    {\it Key words and phrases.} #1}}

\def\ps@firstpage{\ps@empty \def\@oddhead{\hss\pplogo}%
  \let\@evenhead\@oddhead 
}
\def\maketitle{\par
 \begingroup
 \def\thefootnote{\fnsymbol{footnote}}
 \def\@makefnmark{\hbox
 to 0pt{$^{\@thefnmark}$\hss}}
 \if@twocolumn
 \twocolumn[\@maketitle]
 \else \newpage
 \global\@topnum\z@ \@maketitle \fi\thispagestyle{firstpage}\@thanks
 \endgroup
 \setcounter{footnote}{0}
 \let\maketitle\relax
 \let\@maketitle\relax
 \gdef\@thanks{}\gdef\@author{}\gdef\@title{}\let\thanks\relax}

\def\abstract{\if@twocolumn
\section*{{\bf Abstract}}
\else \small
\begin{center}
{\bf ABSTRACT}
\end{center}
\quotation
\fi}

\def\thebibliography#1{\section*{References\@mkboth
 {REFERENCES}{REFERENCES}}\small\list
 {[\arabic{enumi}]}{\settowidth\labelwidth{[#1]}\leftmargin\labelwidth
 \advance\leftmargin\labelsep
 \usecounter{enumi}}
 \def\newblock{\hskip .11em plus .33em minus .07em}
 \sloppy\clubpenalty4000\widowpenalty4000
 \sfcode`\.=1000\relax}

\newif\iffn\fnfalse

\@ifundefined{reset@font}{\let\reset@font\empty}{} 
\long\def\@footnotetext#1{\insert\footins{\reset@font\footnotesize
    \interlinepenalty\interfootnotelinepenalty
    \splittopskip\footnotesep
    \splitmaxdepth \dp\strutbox \floatingpenalty \@MM
    \hsize\columnwidth \@parboxrestore
   \edef\@currentlabel{\csname p@footnote\endcsname\@thefnmark}\@makefntext
    {\rule{\z@}{\footnotesep}\ignorespaces
      \fntrue#1\fnfalse\strut}}}

\makeatother

\begin{document}
\setcounter{page}0
\title{\LARGE {\bf String Loop Threshold Corrections 
 for N=1 Generalized Coxeter Orbifolds}\\[10mm]}
\author{
{\bf C. Kokorelis}\\[0.1cm]
\normalsize CITY University Business School\\
\normalsize Department of Investment, Risk Management and Insurance\\
\normalsize Frobischer Crescent, Barbican Centre,
London, EC2Y 8HB,U.K\\[5mm]
}

{\hfuzz=10cm\maketitle}

\def\Large{\large}
\def\LARGE{\large\bf}

\vskip 1cm

\begin{abstract}
We discuss
the calculation of threshold corrections to gauge coupling constants for the,
 only, non-decomposable 
class of abelian (2,2) symmetric N=1 four dimendional
heterotic orbifold models, where
the internal twist is realized as a generalized Coxeter automorphism.
The latter orbifold was singled out by earlier work as
the only N=1 heterotic $Z_N$ orbifold that satisfy the phenonelogical 
criteria of correct minimal gauge coupling
unification and cancellation of target space modular anomalies. 
\end{abstract}

\newpage

The purpose of this paper is to examine the appearance of one-loop string
threshold
corrections in the gauge couplings of the four dimensional generalized
non-decomposable $N=1$ orbifolds of the
heterotic string.
In 4${\cal D}$ $N=1$ orbifold compactifications
the process of integrating out 
massive string modes, causes
the perturbative one-loop threshold corrections\footnote{which
receive non-zero moduli dependent
corrections from the $N=2$ unrotated sectors.},
to receive non-zero corrections in the form
of automorphic functions of the target space modular group.
At special points in the moduli space previously massive states
become
massless and contribute to gauge symmetry enhancement. As a result the
appearance of massless
states in the running coupling constants appears in the form of a dominant
logarithmic term \cite{cve,kokos}.

The moduli dependent threshold corrections of the $N=1$ 4D orbifolds
receive non-zero one loop corrections from orbifold sectors for which
there is a complex plane of the torus $T^6$ left fixed by the orbifold
twist $\Theta$. When the $T^6$ can be decomposed into the direct sum
$T^2 \oplus T^4$, the one-loop moduli dependent threshold 
corrections (MDGTC)
are invariant under the $SL(2,Z)$ modular group and are
classified as decomposable. Otherwise, when the
action of the lattice twist on the $T_6$ torus does not decompose into the
orthogonal sum $T_6 = T_2 \oplus T_4$ with the fixed plane lying
on the $T_2$ torus, MDGTC are invariant under
subgroups of $SL(2,Z)$ and the associated orbifolds are called
non-decomposable.
The $N=1$ perturbative decomposable MDGTC have been calculated,
with the use of string amplitudes, in
\cite{dkl}. The one-loop MDGTC
integration technique of \cite{dkl} was extended
to non-decomposable orbifolds in
\cite{mastie1}. Further calculations of non-decomposable orbifolds
involved in the classification list of $N=1$ orbifolds of \cite{erkl}
have been performed  in \cite{thom1}.

Here we will perform the calculation of one-loop
threshold corrections
for the class of $Z_8$ orbifolds, that can be found
in the classification list of \cite{erkl},
 defined by the Coxeter twist, $\Theta = exp[\frac{2 \pi i}{8}(1, -3, 2)]$ on the root lattice
of $A_3 \times A_3$.
This orbifold was missing from the list of calculations of MDGTG 
of non-decomposable orbifolds
of \cite{mastie1,thom1}
and consequently its one-loop
moduli dependent gauge coupling
threshold corrections 
were not calculated in \cite{mastie1, thom1}.
In \cite{kokos1,kokos2} we found that this orbifold is non-decomposable
and it is the only one that poccesses this property
from the list of generalized Coxeter orbifolds
given in \cite{erkl}.
Its twist can be equivalently realized
through the generalized Coxeter automorphism
$S_1 S_2 S_3 P_{35} P_{36} P_{45}$ on the root lattice.

Moreover in \cite{kokos2}, where a classification list
of the non-perturbative
gaugino condensation
generated superpotentials and $\mu$-terms of all the $N=1$ four dimensional non-decomposable 
heterotic orbifolds was calculated, its non-perturbative gaugino condensation
generated superpotential was given.
In this work we will calculate its MDGTC following the technique
of \cite{mastie1}.
Our calculation completes the calculation of the
threshold corrections for the
classification list of four dimensional Coxeter
orbifold compactifications with $N=1$
 supersymmetry of \cite{erkl}.

The generalized
Coxeter automorphism is defined as a product of the Weyl reflections%
\footnote{%
The Weyl reflection $S_{i} $ is defined as a reflection 
\begin{equation}
S_{i}(x) = x - 2 {\frac{<x,e_{i}>}{<e_{i},e_{i}>}}e_{i},  \label{coxe1}
\end{equation}
with respect to the hyperlane perpendicular to the simple root.} $S_{i} $ of
the simple roots and the outer\footnote{an automorphism is called outer
if it cannot be generated by a Weyl reflection.}
automorphisms, the latter represented by the
transposition of the roots. An outer automorphism represented by a
transposition which exchanges the 
roots $i \leftrightarrow j$, is denoted by 
$P_{ij}$ and is a symmetry of the Dynkin diagram.

In string theories the one-loop gauge couplings below the string scale 
evolves according to the RG equation
\begin{equation}
\frac{1}{g_a^{2}(p^2)} = \frac{k_a}{g^2_{M_{string}}} +
\frac{b_a}{16 \pi^2}\ln \frac{M_{string}^2}{p^2}+ \frac{1}{16 \pi^2}
 \triangle_a,
\label{rga}
\end{equation}
where $M_{string}$ the string scale
and $b_a$ the $\beta$-function coefficient of all orbifold sectors.
For decomposable orbifolds \cite{dkl} the MDGTC ${\triangle}_a$ 
associated with the gauge couplings $g_a^{-2}$ corresponding to the
gauge group $G_a$, 
are determined in terms of  
the $N=2$ sectors, fixed under both (g, h) boundary conditions, 
of the orbifold, namely
\begin{equation}
\triangle = \int_{\cal F} \frac{d^2 \tau}{\tau_2} \sum_{(g,h)} b_a^{(h, g)}
{\cal Z}_{(h, g)}(\tau, {\bar \tau})- b_a^{N=2}
\int_{\cal F} \frac{d^2 \tau}{\tau_2}.
\label{asdroufe1}
\end{equation}
Here, $b_a^{N=2}$ is the $\beta$-function coefficient of all the $N=2$ 
sectors of
the orbifold, $b_a^{(h, g)}$, $Z_{(h,g)}$ the $\beta$-function coefficient
of the $N=2$
sector untwisted under $(g, h)$ and its partition function (PF)
respectively.
The integration is over the
fundamental domain $\cal F$ of the $PSL(2, Z)$.
For the case of non-decomposable orbifolds the situation is slightly
different, namely   
\begin{equation}
\triangle = \int_{{\tilde {\cal F}}} \frac{d^2 \tau}{\tau_2} \sum_{(g_0,h_0) \in
{\cal O}} b_a^{(h_0, g_0)}
{\cal Z}_{(h_0, g_0)}(\tau, {\bar \tau})- b_a^{N=2}
\int_{\cal F} \frac{d^2 \tau}{\tau_2}.
\label{asdroufe2}
\end{equation}
The difference with the decomposable case now is 
that the sum, in the first integral of (\ref{asdroufe2}) is
over those $N=2$ sectors
that belong to the $N=2$ fundamental orbit $\cal O$ and the 
integration is not over $\cal F$ but over the fundamental
domain $\tilde {\cal F}$. 
Because the PF ${\cal Z}_{(g_0, h_0)}$,  for non-decomposable orbifolds,
is invariant under
subgroups of the modular group i.e  $\tilde \Gamma$, the domain
${\tilde {\cal F}}$
is generated by the action of those  modular transformations 
that generate $\tilde \Gamma$ from $\Gamma$. In the example
that we examine later in this work, $\tilde \Gamma = \Gamma_o(2)$ and
${\tilde {\cal F}} = \{1, S, ST\}{\cal F}$.
In turn the fundamental orbit $\cal O$ is generated by the action of
$\tilde {\cal F}$ on the fundamental element of this orbit.

For the orbifold ${Z_8}$ there are four complex moduli fields.
There are three $(1,1)$ moduli due to the three untwisted generations $27$
and one $(2,1)$-modulus due to the one untwisted generation ${\bar 27}$.
The realization of the point group is generated by 
\begin{equation}
Q=\left( 
\begin{array}{cccccc}
0 & 0 & 0 & 0 & 0 & -1  \\ 
1 & 0 & 0 & 0 & 0 & 0  \\ 
0 & 1 & 0 & 0 & 0 & -1 \\ 
0 & 0 & 1 & 0 & 0 & 0  \\ 
0 & 0 & 0 & 1 & 0 & -1 \\ 
0 & 0 & 0 & 0 & 1 & 0  
\end{array}
\right).
\end{equation}
If the action of the
generator of the point group leaves some complex plane invariant then the
corresponding threshold corrections have to depend on the associated moduli
of the unrotated complex plane.
There are three complex untwisted moduli: three $(1,1)$--moduli and no $%
(2,1) $-modulus due to the three untwisted ${\bf {27}}$ generations and 
non-existent untwisted ${\bf \bar 27}$ generation.
The metric $g$ (defined by $g_{ij}=<e_i|e_j>$) has three and the antisymmetric tensor
field $B$ an other three real deformations. The equations $gQ=Q^{\ast}g$
and\footnote{%
By definition $(){^*}$ mean $((){^T})^{-1}$.}
$bQ=Q^{\ast}b$
determine the background fields in
terms of the independent deformation parameters.
\newline
Solving the background field equations one obtains for the metric
\begin{equation}
G=\left( 
\begin{array}{llccrr}
R^2 & u & v & -u & -2v-R^2 & -u \\ 
u & R^2 & u & v & -u & -2v-R^2 \\ 
v & u & R^2 & u & v & -u \\ 
-u & v & u & R^2 & u & v \\ 
-2v-R^2 & -u & v & u & R^2 & u \\ 
-u & -2v-R^2 & -u & v & u & R^2
\end{array}
\right),
\end{equation}
 with $R,u,v \in \Re$ and the antisymmetric tensor field : 
\begin{equation}
B=\left( 
\begin{array}{cccccccc}
0 & x & z & y & 0 & -y   \\ 
-x & 0 & x & z & y & 0  \\ 
-z & -x & 0 & x & z & y \\ 
-y & -z & -x & 0 & x & z   \\ 
0 & -y & -z & -x & 0 & x   \\ 
y & 0 & -y & -z & -x & 0  
\end{array}
\right),
\end{equation}
with $x,y, z \in \Re $.

 The N=2 orbit is given by these sectors which
contain completely unrotated planes, ${\cal O} ={(1,\Theta^4),(\Theta^4,1),(%
\Theta^4,\Theta^4)}$.

The element $(\Theta^4,1)$ can be obtained from the fundamental element
$(1,\Theta^4)$ by an $S$--transformation on $\tau$ and similarly $%
(\Theta^4,\Theta^4)$ by an $ST$--transformation. The partition function for
the zero mode parts $Z_{(g,h)}^{torus}$ of the fixed plane takes the
following form\cite{erkl}
\begin{eqnarray}
Z_{(1,\Theta^4)}^{torus}(\tau ,\bar \tau,G,B) &=& \displaystyle %
\mathop{\sum_{P \in( \Lambda_{N^{\bot})}} q^{\frac{1}{2} {P_L}^{2}} \bar
q^{{P_R}^{2}}},  \nonumber \\
Z_{(\Theta^4,1)}^{torus}(\tau ,\bar \tau,G,B) &=& \displaystyle %
\mathop{\frac{1} {V_{\Lambda_{N}^{\bot}}} \sum_{P \in
({\Lambda}_{N}^{\bot})^{\ast}} q^{\frac{1}{2} {P_L}^2} \bar
q^{\frac{1}{2}{P_R}^{2}}},  \nonumber \\
Z_{(\Theta^4,\Theta^4)}^{torus}(\tau ,\bar \tau,G,B)& =&\displaystyle %
\mathop{\frac{1} {V_{\Lambda_{N}^{\bot}}} \sum_{P \in
({\Lambda}_{N}^{\bot})^{\ast}} q^{\frac{1}{2} {P_L}^2} \bar
q^{\frac{1}{2}{P_R}^{2}}} q^{i \pi(P^2_L - P^2_R)},
\end{eqnarray}
where with $\Lambda_{N}^{\bot}$ we denote the Narain lattice of $A_3 \times A_3$
which has momentum vectors
\begin{equation}
P_L = \displaystyle{\frac{p}{2}+(G-B)w\ ,} \hspace{1cm} P_R = \displaystyle{%
\frac{p}{2}-(G+B)w\ }  \label{zero}
\end{equation}
and $\Lambda_{N^{\bot}}$ is that part of the lattice which remains fixed under $%
Q^4$ and $V_{\Lambda^\bot_N}$ its volume. The lattice
in our case is not self dual in contrast with the case of partition
functions $Z_{(g,h)}^{torus}(\tau,\bar \tau,g,b)$ of \cite{dkl}. Stated
differently the general result is - for the case of non-decomposa\newline
ble orbifolds - that the modular symmetry group is some subgroup of $\Gamma$
and as a consequence the partition function $\tau_2Z_{(g,h)}^{torus}
(\tau,\bar \tau,g,b)$ is invariant under the same subgroup of $\Gamma$.

The subspace corresponding to the lattice $\Lambda^\bot_N$ can be described
by the following winding and momentum vectors, respectively:
\begin{equation}
w=\left( 
\begin{array}{c}
n^1 \\ 
n^2 \\ 
0 \\ 
0 \\ 
n^1 \\ 
n^2
\end{array}
\right)\ ,\ \ n^1,n^2 \in Z\ \ \ {\rm and}\ \ \ p= \left( 
\begin{array}{c}
m_1 \\ 
m_2 \\ 
-m_1 \\ 
-m_2 \\ 
m_1 \\ 
m_2
\end{array}
\right) \ , m_1,m_2 \in Z.  \label{vecz4}
\end{equation}

They are determined by the equations $Q^4w=w$ and $Q^{\ast^4}p=p$ . The
partition function $\tau_2 Z_{(1,\Theta^4)}^{torus}(\tau,\bar \tau,g,b)$ is
invariant under the group $\Gamma_0(2)$, congruence
subgroup of $\Gamma$.  Before we discuss the calculation of threshold
corrections let us give some details about congruence subgroups.
The homogeneous modular group $\Gamma^{\prime}\equiv SL(2,Z)$ is defined as
the group of two by two matrices whose entries are all integers and the
determinant is one. It is called the "full modular group and we symbolize it
by $\Gamma^{\prime}$. If the above action is accompanied with the quotient ${%
\Gamma} \equiv PSL(2,Z) \equiv \Gamma^{\prime}/{\{ \pm 1 \} }$ then this is
called the 'inhomogeneous modular group' and we symbolize it by $\Gamma$.
The fundamental domain of $\Gamma$ is defined as the set of points
which are related through linear transformations $\tau \rightarrow \frac{%
a\tau+b}{c\tau+d}$. If we denote $\tau = \tau_{1} + \tau_{2}$ then the
fundamental domain of $\Gamma$ is defined through the relation ${\cal F} =
\{ \tau \in C |\tau_{2} > 0,|\tau_{1}| \leq \frac{1}{2}, |\tau| \geq 1 \}$.
One of the congruence subgroup of the modular group $\Gamma$ is the group $%
\Gamma_0(n)$.
The group $\Gamma_0(2)$ can be represented by the following set of matrices
acting on $\tau$ as $\tau \rightarrow \frac{a\tau+b}{c\tau+d}$:

\begin{equation}
\Gamma_0(2)=\{ \left( 
\begin{array}{lr}
a & b \\ 
c & d
\end{array}
\right)\; |\; ad-bc=1, \left( c=0\; mod\; 2 \right)\}  \label{groupe}
\end{equation}
It is generated by
the elements $T$ and $ST^2S$ of $\Gamma$. Its fundamental domain is
different from the group $\Gamma$ and is represented from the coset
decomposition $\tilde {{\cal F}}=\{1,S,ST\}{\cal F}$. In addition the group
has cusps at the set of points $\{\infty, 0\}$. Note that
 the subgroup $\Gamma^0(2)$ of $SL(2,
{\bf Z)}$ is defined as with $b = 0\ \bmod 2$.

The integration of the contribution of the various sectors $(g,h)$ is over
the fundamental domain for the group $\Gamma_{0}(2)$ which is a three fold
covering of the upper complex plane. 
By taking into account the values of the momentum and winding vectors in the
fixed directions we get for $Z_{(1,\Theta^4)}^{torus}$
\begin{eqnarray}  \label{general}
Z_{(1,\Theta^4)}^{torus}(\tau,\bar \tau,g,b)= {\sum_{(P_L,P_R) \in {%
\Lambda_{N}^{\bot}}} } q^{\frac{1}{2} P_L^t G^{-1} P_L} q^{\frac{1}{2} P_R^t
G^{-1} P_R} &  \nonumber \\
={\sum_{p,w} e^{2 \pi i \tau p^t w}} e^{-\pi \tau_2 (\frac{1}{2}p^t G^{-1}p
- 2p^t G^{-1} B w + 2 w^t G w - 2 w^t B G^{-1} B w - 2 p^tw)}.& 
\end{eqnarray}
Consider now the the following parametrization of the torus $T^{2}$,
namely define the the $(1,1)$ $T$ modulus and the $(2,1)$ $U$ modulus as:
\begin{equation}
\begin{array}{ccccl}
T & = & T_1+iT_2 & = & 2(b+i \sqrt{\det g_\bot}), \\ 
U & = & U_1+iU_2 & = & \frac{1}{G_{\bot 11}}(G_{\bot 12}+\sqrt{\det G_\bot}),
\end{array}
\end{equation}
where $g_\bot$ is uniquely determined by ${w^t} G w
= (n^1 n^2) {G_\bot} \left(%
{{n^1 }{n^2 }}\right)$ and
b the value of the $B_{12}$ element of the two-dimensional matrix B
of the antisymmetric field. This way one gets
\begin{eqnarray}
T & = & 4(x - y) + i\; 8\; v, \\
U & = & {i}.  \label{comple1}
\end{eqnarray}
Even if we have said that we expect that this $Z_8$ orbifold doe not have
a $h^{(2, 1)}$ U-modulus field, the $T^2$ torus has a U-modulus. 
However 
its value for the $Z_8$ orbifold is fixed. 
The partition function $Z_{(1,\Theta^4)}^{torus}(\tau,\bar \tau,g,b)$ takes
now the form
\begin{equation}
Z_{(1,\Theta^4)}^{torus}(\tau,\bar \tau,T,U)= \sum_{{m_1,m_2 \in 2Z };\;{
n^1,n^2 \in Z}} e^{2 \pi i \tau (m_1 n^1 + m_2 n^2)} e^{\frac{-\pi \tau_2}{%
T_2U_2} |T U n^2+T n^1-Um_1+m_2|^2}\ .  \label{partition}
\end{equation}

By Poisson resummation on $m_1$ and $m_2$, using the identity:
\begin{equation}
\displaystyle\mathop{ \sum_{p \in \Lambda \ast} e^{[-
\pi(p+\delta)^{t}C(p+\delta)] +2\pi i p^t
\phi]}={V}_{\Lambda}^{-1}{\frac{1}{\sqrt det{C}} {\sum}_{l\in
\Lambda}e^{[-\pi(l+\phi)^{t} C^{-1}(l+\phi) - 2 \pi i {\delta}^t
(l+\phi)]}},\;\;\;\;\;\;\;\;\;\;\;\;}  \label{karo}
\end{equation}
we conclude
\begin{equation}
\begin{array}{lrlcr}
& \tau_2 \ Z_{(1,\Theta^4)}^{torus}(\tau,\bar \tau,T,U) & = & {{\frac{1}{4}} 
{\sum_{A \in {\cal M}}} e^{- 2 \pi i T \det A}\ T_2 e^{\frac{-\pi T_2}{%
\tau_2 U_2} \left|(1,U)A \left({\tau }{1}\right) \right|^2}\ ,} &
\label{simrto}
\end{array}
\end{equation}
where 
\begin{eqnarray}
{\cal M}& =& \left( 
\begin{array}{cc}
n_1 & \frac{1}{2} l_1 \\[1mm] 
n_2 & \frac{1}{2} \l_2
\end{array}
\right)
\end{eqnarray}
and $n_1,n_2, l_1,l_2 \in Z$.

From (\ref{simrto}) one can obtain $\tau_2Z_{(\Theta^4,1)}^{torus} (\tau,\bar \tau)$
by an $S$--transformation on $\tau$. After exchanging $n_i$ and $l_i$ and
performing again a Poisson resummation on $l_i$ one obtains

\begin{equation}
\displaystyle\mathop{ Z_{(\Theta^4,1)}^{torus}(\tau,\bar \tau,T,U) =
{\frac{1}{4}} {\sum_{\stackrel{m_1,m_2 \in Z}{ n^1,n^2 \in Z}}} e^{2 \pi i
\tau (m_1 \frac{n^1}{2} + m_2 \frac{n^2}{2})}
e^{{\frac{-\pi\tau_2}{T_2U_2}}|T U{\frac{n^2}{2}}+T{\frac{n^1}{2}}- U m_1 +
m_2|^2}}.  \label{pa}
\end{equation}
The factor $4$ is identified with the volume of the invariant sublattice in
(\ref{pa}). The expression $\tau_2 Z_{(\Theta^4,1)}^{torus}(\tau,\bar \tau,T,U)$
is invariant under $\Gamma^0(2)$ acting on $\tau$ and is identical to that
for the $(\Theta^4,\Theta^4)$ sector.

Thus
the contribution of the two sectors $(\Theta^4,1)$ and $(\Theta^4,\Theta^4)$
to the coefficient $b_a^{N=2}$ of the $\beta$--function is one fourth of
that of the sector $(1,\Theta^4)$, thus
\begin{equation}
b^{N=2}_a = \frac{3}{2} b_a^{(1,\Theta^4)}.
\label{daq12}
\end{equation}
The coefficient $b^{N=2}_a$ is the contribution to the $\beta$ functions of
the $N=2$ orbit.
Including the moduli dependence of the different sectors, we conclude that
the final result for the threshold correction to the inverse gauge coupling
 reads
\begin{equation}  
{\triangle_{a}(T,\bar T,U,\bar U)} = -b_a^{N=2} \cdot  \ln|{\frac{8\pi
e^{1-\gamma_E}}{3 \sqrt{3}} } T_2 |\eta\left({\frac{T}{2}}%
\right)|^4 U_2 |\eta(\left( U \right)|^4 |.
\label{rez4}
\end{equation}
The value of $U_2$ is fixed and equal to one as can be easily seen from
eqn.(\ref{comple1}). In general for $Z_N$ orbifolds with $N \geq 2$ the value of
the U modulus is fixed. The final duality symmetry of (\ref{rez4}) is ${%
\Gamma^{0}_{T}(2)}$ with the value of U replaced with the
constant value i.

Let us use (\ref{asdroufe2}), (\ref{rez4}) to deduce
some information about
the phenomenology of the $Z_8$ orbifold considered in this work.
We want to calculate the 
one-loop corrected string mass unification scale $M_X$, that is when two gauge
group
coupling constants become equal, i.e. 
$\frac{1}{k_a g_a^2}= \frac{1}{k_b g_b^2}$. We further assume that
the gauge group of our theory at the string unification scale if
given by $G = \oplus G_i$, where $G_i$ a gauge group factor.
Taking into account (\ref{rez4}) we get 
\begin{eqnarray}
M_X = M_{string} 
[T_2 |\eta(\frac{T}{2})|^4 U_2 
|\eta(U)|^4]^{\frac{(b_b^{N=2} k_a -b_a^{N=2} k_b)}{2(b_a k_b - b_b k_a)}},\nonumber\\
M_{string} \approx 0.7 g_{string} 10^{18}\; Gev.
\label{unif}
\end{eqnarray}
where $k_i$ the Kac-Moody level associated to the gauge group factor $G_i$.

We will give now some details about the integration in ({\ref{asdroufe2})
of the integral that we
used so far to derive (\ref{rez4}). The integration of eqn. (\ref{partition}) is
over a $\Gamma_0(2)$
subgroup of the modular group $\Gamma$ since (\ref{partition}) is invariant
under a $\Gamma_0(2)$ transformation $\tau \rightarrow
\frac{a\tau+b}{c\tau+d}$ (with $ad-bc=1, c=0 \bmod 2$).
Under a $\Gamma_0(2)$
transformation (\ref
{partition}) remains invariant if at the same time we redefine our
integers $n_1,n_2,l_1$ and $l_2$ as follows:
\begin{equation}
\left( 
\begin{array}{cc}
n^{\prime}_1 & n^{\prime}_2 \\ 
l^{\prime}_1 & l^{\prime}_2
\end{array}
\right)=\left( 
\begin{array}{cc}
a & c/2 \\ 
2b & d
\end{array}
\right) \left( 
\begin{array}{cc}
n_1 & n_2 \\ 
l_1 & l_2
\end{array}
\right)  \label{trans}
\end{equation}
The integral can be calculated based on the method of decomposition into
modular orbits. There are three set's of inequivalent orbits under the $%
\Gamma_0(2)$, namely : 
\newline
$a.$) {\em The degenerate orbit of zero matrices}, where after integration over $%
\tilde {{\cal F}}=\{1,S,ST\}{\cal F}$ gives as a total contribution $I_{0} =
\pi T_2 /4 $. 
\newline
$b.$) {\em The orbit of matrices with non-zero determinants}. The following
representatives give a non-zero contribution $I_1$ to the integral:
\begin{equation}
\left( 
\begin{array}{cc}
k & j \\ 
0 & p
\end{array}
\right)\ ,\ \left( 
\begin{array}{cc}
0 & -p \\ 
k & j
\end{array}
\right)\ ,\ \left( 
\begin{array}{cc}
0 & -p \\ 
k & j+p
\end{array}
\right)\ ,\ \ 0\leq j < k , \ \ p \neq 0\ ,  \label{i2m}
\end{equation}
where $ I_o + I_1 =- (3/2) \cdot 4  Re \ln\eta(\frac{T}{2})$
\newline
$c.$) {\em The orbits of matrices with zero determinant},
\begin{equation}
\left( 
\begin{array}{cc}
0 & 0 \\ 
j & p
\end{array}
\right)\ ,\ \left( 
\begin{array}{cc}
j & p \\ 
0 & 0
\end{array}
\right)\ ,\ j,p \in Z\ ,\ (j,p) \neq (0,0)\ .  \label{rouk}
\end{equation}
The first matrix in (\ref{rouk}) has to be integrated over the half--band $%
\{\tau \in C \; \tau_2>0\; ,\;|\tau_1|<\;h\}$ while the second
matrix has to be integrated over a half--band
with the double width in $\tau_1$.
The total contribution from the modular orbit $I_3$ gives,
\[
\begin{array}{ll}
I_3= & \displaystyle{-4 {\Re} \ln \eta(U)-\ln \left(T_2 U_2\right)
+ \left(\gamma_E-1-\ln \frac{8\pi}{3\sqrt{3}}\right)} \\ 
& \displaystyle{-\frac{1}{2} \times 4 {\rm Re} \ln \eta(U)-\frac{1}{2}
\times \ln \left(T_2U_2\right)+\frac{1}{2} \times (\gamma_E-1-\ln \frac{8\pi%
} {3\sqrt{3}})}.
\end{array}
\]
Putting $I_o$, $I_I$, $ I_2$ together we get (\ref{rez4}).

All $N=1$ four dimensional orbifolds have been tested in \cite{ibalu}
as to whether they satisfy several phenomenological criteria, involving
\newline
a) correct unification of the three gauge coupling constants 
at a scale $M_X \approx 10^{16}$ Gev, assuming 
the minimal supersymmetric, Standard Model gauge group $G= SU(3) \times 
SU(2) \times U(1)$, particle spectrum with a SUSY threshold close 
to weak scale, in the two cases of i) a single overall modulus in the three 
complex planes $T= T_1 = T_2 = T_3$ and ii) the anisotropic squeezing case 
$T_1 >> T_2, T_3$,
\newline
\newline
b) anomaly cancellation with respect to duality transformations
of the moduli in the planes rotated by all the orbifold twists.
\newline
\newline
The only orbifold from this study that satisfy all the
phenonemenological criteria, set out by Ib\'{a}$\tilde{n}$ez and L$\ddot{u}$st,
is the $Z_8$ orbifold that we examined in this work.

\end{document}